\documentclass[12pt,a4paper,reprint,aip,pof,onecolumn]{revtex4-1}
\usepackage{amsmath,amssymb,amsthm}
\usepackage{graphicx, epsfig, upgreek}
\usepackage{natbib}
\usepackage{hyperref}
\usepackage[left=1.in,right=1.in,top=1in,bottom=1in]{geometry}

\begin{document}
\title{Effect of Electro-Osmotic Flow on Energy Conversion on Superhydrophobic Surfaces}

\author{Gowrishankar Seshadri}
\affiliation{Department of Chemical Engineering, Indian Institute of Technology, Bombay, Powai, Mumbai - 400076, India}

\author{Tobias Baier}
\email[Corresponding author: ]{baier@csi.tu-darmstadt.de}
\affiliation{Center of Smart Interfaces, Petersenstr. 17, 64287 Darmstadt, Germany}

\date{\today}

{\noindent \small \textit{The following article has been accepted by Physics of Fluids. After it is published, it will be found at http://pof.aip.org/.}}

\begin{abstract}
It has been suggested that superhydrophobic surfaces, due to the presence of a no-shear zone, can greatly enhance transport of surface charges, leading to a considerable increase in the streaming potential. This could find potential use in micro-energy harvesting devices. In this paper, we show using analytical and numerical methods, that when a streaming potential is generated in such superhydrophobic geometries, the reverse electro-osmotic flow and hence current generated by this, is significant. A decrease in streaming potential compared to what was earlier predicted is expected. We also show that, due to the electro-osmotic streaming-current, a saturation in both the power extracted and efficiency of energy conversion is achieved in such systems for large values of the free surface charge densities. Nevertheless, under realistic conditions, such microstructured devices with superhydrophobic surfaces have the potential to even reach energy conversion efficiencies only achieved in nanostructured devices so far.
\end{abstract}

\maketitle


\renewcommand {\baselinestretch} {1.1} \normalsize
\fontsize{11pt}{15pt}\selectfont

\section{Introduction}
\label{sec:intro}

Energy conversion and flow detection devices, fabricated in the micro- and nano-scale, have a potential application in novel highly integrated or portable systems. One way to achieve conversion from hydraulic pressure to electric energy is via the streaming current. The phenomenon was first reported by Quincke\cite{quincke} in 1859, who noted that when pure water flows trough a porous material an electric current is generated. Helmholtz\cite{helmholtz}, expanding on Quincke's reasoning, gave the first theoretical explanation for this, noting that virtually any surface brought in contact with water will acquire a net charge, attached to the surface, that is screened in a diffuse charge cloud within the water in the vicinity of the surface. Thus when fluid flows through a capillary, the mobile part of this electrical double layer is dragged along, leading to the streaming current. However, since the electric double layer naturally arises at solid walls where the flow velocity is small, conversion rates are usually low \cite{probstein, li, zhao11}, reaching a few percent in channels of nanometer dimension\cite{vdHeyden_2007, vdHeyden_2006} and much less in microchannels. It was therefore suggested that an effective wall-slip can increase the efficiency of such a device\cite{yang_2003, yang_2004, davidson, Ren_2008, chakraborty}.

Some superhydrophobic surfaces trap pockets of air within grooves. This so-called Cassie-Baxter state has attracted much attention due to the reduced drag experienced by a fluid flowing over such a surface \cite{ybert, rothstein}. In particular shear and pressure driven flow over a striped geometry has been thoroughly studied\cite{philip, lauga, prosperetti, Bazant_2008, Huang_2008}. Since the air-water interface can be considered as posing no resistance to shear, in principle such a surface seems ideal for an increase in streaming current, as long as the free surface carries a net charge. This latter condition seems nontrivial at first, since any ions that are preferentially attracted to the surface are expected to attract an equal amount of opposite charge in a diffuse cloud just as on a solid surface. However, a net charge is attracted to the surface when an electric field is applied normal to the fluid-air interface\cite{melcher}, for example by embedding electrodes within the grooves\cite{steffes}.

It has been shown that the superhydrophobic nature of the surface, together with a charged air-water interface, can be exploited in order to generate a significant increase in streaming potential\cite{zhao11}. However, in this study the effect of electro-osmotic (EO) flow due to the built-up streaming potential was neglected, which generally decreases the efficiency since it goes in the opposite direction as the pressure driven flow\cite{osterle}. Moreover, on superhydrophobic surfaces the EO flow is enhanced by essentially the same mechanism and magnitude as the streaming current, namely due to charges on the no-shear surface\cite{squires, bazant1, Belyaev_2011}. In this paper therefore, we show that on superhydrophobic surfaces, the electrokinetic flow resulting from the streaming potential, together with its streaming current, plays an important role in limiting the efficiency of energy harvesting with increasing free surface charge and effective slip length. So far, the only paper where both effects, in particular the EO streaming current, have been properly taken into account for energy conversion on patterned surfaces seems to be Ng and Chu's work\cite{Ng_2011}. Their focus, however, is on the EO flow and the streaming current due to pressure driven flow, only briefly exploring energy conversion on a patterned surface with alternating no-slip and finite-slip stripes. Here, our attention is on energy conversion in the optimal case of alternating slip and no-slip surfaces, a good approximation to a liquid in Cassie-Baxter state above air-filled cavities. Further, in our analysis we numerically go beyond the Debye-H\"uckel approximation and explore its impact on the results for energy conversion.

We organize this paper into three sections. In section~\ref{sec:generalformulations} we present the general formulations and derive expressions for the pressure driven flow, the EO flow and energy conversion and harvesting. In section~\ref{sec:resultsanddiscussion}, we present the results of our study in two parts. We first discuss our results through analytical approximations of the expressions derived in section~\ref{sec:generalformulations}, and then graphically present the exact behavior of the expressions presented in the section~\ref{sec:generalformulations}, and validate our analytical results. In section~\ref{sec:conclusion}, we present our conclusions based on this study.

\section{General Formulations}\label{sec:generalformulations}
\begin{figure}[!tb]
\includegraphics[width=9 cm]{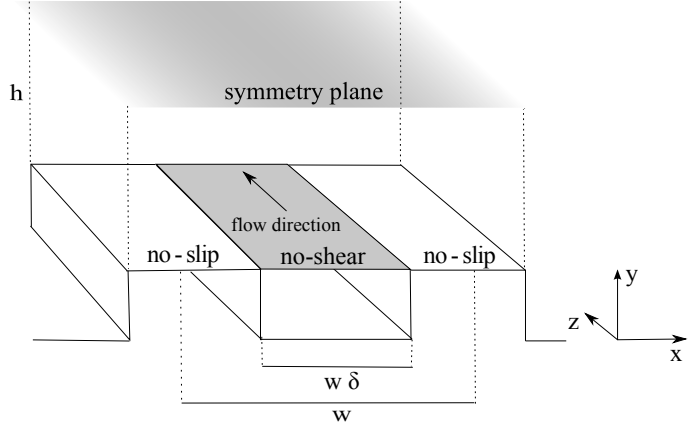}
\caption{Schematic diagram of the superhydrophobic surface. Within the grooves air is entrapped and we assume the air-liquid interface to remain ideally flat. The length of the channel is assumed to be $l$ and consists of two parallel surfaces at distance $2h$. The surfaces are periodically structured with groves of width $w \delta$ and periodicity $w$.}
\label{fig:schematic}
\end{figure}

We consider flow over a striped surface with alternating no-shear and no-slip regions as shown in figure~\ref{fig:schematic}. The liquid-air interface is assumed to be ideally flat, such that the liquid occupies the domain $y\geq 0$. Zhao \cite{zhao11} showed that the overall efficiency for generation of streaming potential is greater for flows in the longitudinal direction when compared to the transverse direction. Therefore, we here concentrate on flow in the longitudinal direction. The electric potential on the no-slip part of the superhydrophobic surface as well as on the no-shear part is assumed to be constant, denoted by ${\zeta_{ns}}$ and ${\zeta_s}$, respectively. The striped surface is assumed to have a period of $w$ and the height of the channel is taken to be $2h$, with the channel being symmetric about the center line. Since the device is assumed to be in the $\upmu$m range, and the typical Debye screening length is of the order of nm, we make the assumption that ($h \gg \lambda_D$) where $\lambda_D$ is the Debye layer thickness and $\kappa=1/\lambda_D$. This condition will be relaxed later in numerical calculations. The liquid viscosity is denoted by $\mu$ and the permittivity of the liquid is denoted by $\varepsilon$. The free surface fraction, i.e. the ratio of the width of the no-shear region to the total width of the period of the interface, is denoted by $\delta$. A schematic of the setup is shown in figure~\ref{fig:schematic}.

\subsection{Bi-Layer Potential Distribution} \label{sec:bilayer}
For our (semi-) analytic calculation, the potential distribution in the bi-layer is derived based on the Debye-H\"uckel theory\cite{probstein}. The electric potential in the system obeys the Poisson-Boltzmann distribution. Under the assumption that the electric potential energy of the system is much less than the average thermal energy, $k_BT/e\simeq 26$ mV at room temperature, the governing equation for the potential distribution is
\begin{equation}
\label{eq:governv}
\nabla^2V(x,y)=\kappa^2V(x,y).
\end{equation}
This equation is solved under the boundary conditions that, $V(x,0)=\zeta_{ns}$ for $0<x<\frac{\delta w}{2}$ and $V(x,0)=\zeta_{s}$ for $\frac{\delta w}{2}< x<\frac{w}{2}$, using separation of variables and a Fourier series expansion in the periodic variable $x$. Denoting the average potential on the boundary as $\overline\zeta=\delta\zeta_s+(1-\delta)\zeta_{ns}$, the final form of the solution is given by
\begin{equation}\label{eq:a}
V(x,y)=\overline\zeta e^{-\kappa y}+\sum_{n=1}^\infty \beta_n \cos\left(\omega_n x\right) \exp \left(-\sqrt {\kappa^2+\omega_n^2}\,y \right),
\end{equation}
for $y<h$, where we have used $\omega_n=\frac{2n\pi}{w}$ and $\beta_n=\frac{2[\zeta_s-\zeta_{ns}]\sin n\pi\delta}{n\pi}$. Here and in the following, we will assume $\kappa h\gg 1$, such that exponential functions $\sim \exp(-\kappa y)$ have died away sufficiently in the center of the channel, allowing us to treat the walls independently for all solutions of this type.

Later, we also use a simplified form of equation (\ref{eq:a}), obtained by solving equation (\ref{eq:governv}) independently in the no-shear region and the no-slip region. The resulting solution is defined in the form of a split function as
\begin{equation}\label{eq:vsplit}
\begin{split}
V(x,y) =\zeta_se^{-\kappa y}, &\qquad      0< x< \delta w/2,\\
V(x,y) =\zeta_{ns}e^{-\kappa y},&\qquad  \delta w/2<x<w/2.\\
\end{split}
\end{equation}
This approximates equation (\ref{eq:a}) in a region extending $\mathcal{O}(\lambda_D)$ from the wall. We will comment on the applicability of this approximation in section \ref{sec:eom} where it is used.

The charge-density corresponding to the potential distribution is frequently used and is given by Gauss's law
\begin{equation}
\label {eq:chargedensity}
\rho_e=-\varepsilon\nabla^2V=-\varepsilon \kappa^2V.
\end{equation}

\subsection{Streaming Velocity and Current}
\label{sec:streaming}
The streaming current is generated by pressure-driven flow. Since flow velocities are small, inertial terms can be neglected in the momentum equation and the governing equation for the streaming velocity thus is the Stokes equation
\begin{equation}\label{eq:upEq}
\nabla^2 u_{p}(x,y)= \frac{\Delta P}{\mu l},
\end{equation}
where $\Delta P$ is the (negative) pressure drop applied across the channel of length $l$. This equation is solved for the velocity profile by splitting the velocity into two components, $u_p=u_p^1+u_p^2$. Here, $u_p^1$ is the velocity profile for flow between parallel no-slip plates. The correction, $u_p^2$, to this velocity is then obtained by demanding that $u_p$ obeys the no shear boundary condition on the slipping portion. The flat plate velocity profile is given by
\begin{equation}\label{eq:ustreamflat}
u_p^1(y)=-\frac{\Delta Ph^2}{2 \mu l} \left[\frac{2y}{h}-\left(\frac{y}{h}\right)^2\right].
\end{equation}
The correction to this velocity profile satisfies $\nabla^2u_p^2=0$. Since the geometry is periodic, the velocity profile can be expanded in a Fourier series as
\begin{equation}\label{eq:generalform}
u_p^2(x,y)=a_0+\sum_{n=1}^\infty a_n\cos(\omega_n x)\, e^{-\omega_n y}.
\end{equation}
The net velocity profile, $u_p$, satisfies the boundary conditions that at $y=0$, $\partial_y u_p=0$ in the no shear region and $u=0$ in the no slip region. The velocity correction, $u_p^2$, thus satisfies the conditions at $y=0$ that $\partial_y u_p^2=-\partial_y u_p^1;\;0<x<\frac{\delta w}{2}$ and $u_p^2=0 ; \; \frac{\delta w}{2}<x<\frac{w}{2}$. Using the expansion (\ref{eq:generalform}) these equations translate to
\begin{equation}
\begin{split}\label{eq:dualSeriesP}
\sum_{n=1}^\infty \omega_n a_n\cos \omega_n x =-\frac{\Delta Ph}{\mu l}, &\qquad 0<x<\frac{\delta w}{2},\\
a_0+\sum_{n=1}^\infty a_n\cos\omega_n x=0, &\qquad \frac{\delta w}{2}<x<\frac{w}{2}.
\end{split}
\end{equation}
Systems of equations of this type were solved by Sneddon \cite{sneddon}, such that \cite{prosperetti}
\begin{equation}\label{eq:ustreamcorrection}
\begin{split}
a_0&=-\frac{\Delta Pwh}{2\mu l}\beta_\parallel,\\
a_n&=-\frac{\Delta Pwh}{2\pi\mu l}\int_0^{\delta\pi}\tan\frac{x}{2}\, \left[P_n(\cos x)+P_{n-1}(\cos x)\right]\;dx,
\end{split}
\end{equation}
where $P_n(x)$ are the $n^\mathrm{th}$ order Legendre polynomials and $\beta_\parallel=\frac{2}{\pi}\log\sec\left(\frac{\pi\delta}{2}\right)$ is the non-dimensionalised effective slip length for pressure driven or Couette flow over such surfaces\citep{philip,lauga}.

The electric streaming current through a cross section spanning half a period generated by the pressure driven flow is given by
\begin{equation}\label{eq:istreamdef}
J_{p}=2\int_0^{h}\int_0^{\frac{w}{2}}\rho_eu_{p}\,dx\,dy.
\end{equation}
Using the expressions derived in equations (\ref{eq:a}), (\ref{eq:ustreamflat}) and (\ref{eq:ustreamcorrection}), the above integral is evaluated to give
\begin{equation}
\label{eq:istreamfull}
\begin{split}
J_{p}&=-\frac{\varepsilon \kappa^2 \Delta Phw\overline{\zeta}}{\mu l}\left[1-\frac{1}{\kappa h}\right]+{\varepsilon \kappa\overline{\zeta}wa_0 }+\frac{\varepsilon \kappa^2}{2}\sum_{n=0}^{\infty}\frac{a_n \beta_n w}{\omega_n+\sqrt{\kappa^2+\omega_n^2}}.
\end{split}
\end{equation}
 Since the typical order of $\lambda_D=1/\kappa$ is much less than the width of the plate, we can approximate the sum in the above expression by
 \begin{equation}\label{eq:istreamsim}
 \frac{\varepsilon \kappa^2}{2}\sum_{n=0}^{\infty}\frac{a_n \beta_n w}{\omega_n+\sqrt{\kappa^2+\omega_n^2}}\simeq \frac{\varepsilon \kappa(\zeta_s-\zeta_{ns})w}{\Delta P}\sum_{n=1}^\infty{a_n\frac{\sin(n\pi\delta)}{n\pi}}.
 \end{equation}
By integrating equation (\ref{eq:dualSeriesP}) from $\frac{\delta w}{2}$ to $\frac{w}{2}$, we can show that $\sum_{n=1}^\infty a_n\frac{\sin (n\pi\delta)}{n\pi}=(1-\delta)a_0$. Substituting this in equation (\ref{eq:istreamsim}) we get a simplified expression for the streaming current
 \begin{equation}\label{eq:istreamsimple}
J_p\simeq wh\left[\frac{w\zeta_s \kappa\varepsilon\beta_\parallel}{2\mu l}+\frac{\varepsilon \overline \zeta}{\mu l}\right]\Delta P.
\end{equation} 
Correspondingly, the volumetric flow rate for the pressure driven flow is obtained as
\begin{equation}\label{eq:qp}
Q_p=2\int_0^h\int_0^{\frac{w}{2}}u_p\,dx\,dy=\frac{h^3w\Delta P}{2\mu l}\left(\frac{2}{3}+a_0\right).
\end{equation}

\subsection{Electro-Osmotic Flow and Current}\label{sec:eom}
An electro-osmotic flow, which thereby drives a current, results from the electric field which is generated due to the pressure driven flow. The governing equation for this EO flow is given by
\begin{equation}\label{eq:ueEq}
\nabla^2u_e=\rho_eE_z.
\end{equation}
We again write the velocity profile, $u_e=u_e^1+u_e^2$, as a superposition of a velocity on a no-slip surface with periodic potential, $u_e^1$, and a correction term which arises because of a no-shear zone, $u_e^2$. Assuming that $\partial_x^2 u_e^1\ll\partial_y^2u_e^1$, correct apart from a small region of width of order $\lambda_D$ around $x=\delta w/2$, we solve for $u_e^1$ using equation (\ref{eq:vsplit}) for the electric potential and get
\begin{equation}
\label{eq:uesplit}
\begin{split}
u_e^1(x,y) =\frac{\varepsilon\zeta_s\Delta\phi}{\mu l}(1-e^{-\kappa y}),\qquad &0<x<\frac{\delta w}{2}, \\
u_e^1(x,y) =\frac{\varepsilon\zeta_{ns}\Delta\phi}{\mu l}(1-e^{-\kappa y}),\qquad &\frac{\delta w}{2}<x<\frac{w}{2}.
\end{split}
\end{equation}
Obviously, this is not a full solution of (\ref{eq:ueEq}), since it has a jump at $x=\delta w/2$, just as (\ref{eq:vsplit}) is not a full solution of (\ref{eq:governv}). However, as we will see, the EO velocity field is generally dominated by $u_e^2$ when charge is present at the free surface, so the error introduced due to this approximation is exceedingly small. Moreover, and more importantly, for thin Debye layers, $\kappa w\gg 1$, the approximation is excellent in the vicinity of the wall, such that we can obtain the correct boundary condition for $u_e^2$ from it. Effectively, this amounts to replacing the diffuse charge cloud with a corresponding surface charge density at the interface, resulting in a jump in shear rate across the interface\cite{steffes}.

The general form of $u_e^2$ can again be expanded as a Fourier series
\begin{equation}\label{eq:uecorrection}
u_e^2(x,y)=A_0+\sum_{n=1}^\infty A_n\cos(\omega_n x)\, e^{-\omega_n y}.
\end{equation}
Since the full EO velocity field, $u_e$, satisfies $\partial_y u_{e}=0$ at $y=0$ in the no shear region and $u_e=0$ in the no slip region, the boundary conditions for $u_e^2$ at $y=0$ are $\partial_y u_e^2=-\partial_y u_e^1;\; 0<x<\frac{\delta w}{2}$ and $u_e^2=0 ; \; \frac{\delta w}{2}<x<\frac{w}{2}$. In particular, the Fourier coefficients of $u_e^2$ must satisfy the conditions
\begin{equation}
\label{eq:a0}
\begin{split}
\sum_{n=1}^\infty \omega_n A_n\cos \omega_n x =\frac{\varepsilon\zeta_s \kappa\Delta\phi}{\mu l},&\qquad 0<x<\frac{\delta w}{2}, \\
A_0+\sum_{n=1}^\infty A_n\cos\omega_n x=0, &\qquad \frac{\delta w}{2}<x<\frac{w}{2},
\end{split}
\end{equation}
completely analogous to equation (\ref{eq:ustreamcorrection}). Solving this double series, we get
\begin{equation}
\begin{split}
&A_0=\frac{\varepsilon\zeta_s \kappa w\Delta\phi}{2\mu l}\beta_\parallel,\\
A_n&=\frac{\varepsilon\zeta_s \kappa w\Delta\phi}{2\pi\mu l}\int_0^{\delta\pi}\tan \frac{x}{2}\,\left[P_n(\cos x)+P_{n-1}(\cos x)\right]\,dx.
\end{split}
\end{equation}
We now see that the homogeneous part of the flow, given by $A_0$, is larger than $u_e^1$ by a factor of $\kappa w\beta_\parallel/2$. Thus as long as $\zeta_s/\zeta_{ns}\ll \beta_\parallel w/\lambda_D$, the approximation going into equation (\ref{eq:uesplit}) plays no role in the bulk velocity field.

The electro-osmotic current, $J_{eom}$, is thus given by
\begin{equation}\label{eq:jeomdef}
J_{eom}=2\int_0^{h}\int_0^{\frac{w}{2}}\rho_eu_{e}\,dx\,dy.
\end{equation}
This integral is evaluated in Appendix \ref{app:eomsimplify}, and the expression for the EO current is given by equation (\ref{ieomapprox}) and is shown below
\begin{equation}
\label{eq:ieomfull}
\begin{split}
J_{eom}=2(-\varepsilon \kappa^2)\Bigg[\frac{\varepsilon\zeta_s^2\Delta\phi\delta w}{4\kappa\mu l}&+\frac{\varepsilon\zeta_{ns}^2\Delta\phi(1-\delta) w}{4\kappa\mu l}+\frac{\overline\zeta w^2\varepsilon\zeta_s\Delta\phi}{4\mu l}\frac{2}{\pi}\log\sec\left(\frac{\delta\pi}{2}\right)\\
&\qquad\qquad\quad +(\zeta_s-\zeta_{ns})\sum_{n=1}^\infty \frac{A_n\sin(n\pi d)}{\omega_n(\omega_n+\sqrt{\kappa^2+\omega_n^2})}\Bigg] 
\end{split}
\end{equation}
Under the assumption that ($\kappa\gg\omega_n$), a simplification to the above expression is carried out, similar to the simplification of the streaming current, and is given in Appendix \ref{app:eomsimplify}. The simplified form of the EO current is given by
\begin{equation}
\label{ieomfinalapprox}
J_{eom}=-\left[\left(\frac{\varepsilon^2 \zeta_s^2 \kappa^2 w^2}{2\mu l}\right)\beta_\parallel\right]\Delta\phi.
\end{equation}
Correspondingly, the flow rate due to the EO velocity field is given by
\begin{equation}\label{eq:qe}
Q_e=2\int_0^h\int_0^{\frac{w}{2}} u_e \,dx\,dy=-wh\left(A_0+\frac{\varepsilon\overline{\zeta}\Delta\phi}{\mu l}\left(1-\frac{1}{\kappa h}\right)\right).
\end{equation}

\subsection{Energy Efficiency and Streaming Potential}\label{sec:energyeff}

In our system, apart from the streaming current and the electro-osmotic current, a bulk conduction current through the system is present which is given on the basis of the bulk conductivity of the liquid, $\sigma$, by
\begin{equation}
\label{eq:iconduction}
J_{cond}=\frac{\sigma wh\Delta\phi}{l},
\end{equation}
valid for homogeneous conductivity, i.e. neglecting variations in carrier density within the double layer, a condition we will relax in our numerical evaluation. Energy is extracted from the system by connecting an external load of resistance $R_{ext}$ in series with the channel. The current running through this load is thus $J_{ext}=\Delta\phi/R_{ext}$ and the power extracted is $P_{ext}=\Delta\phi^2/R_{ext}$. 

Due to the linearity of the system, the expressions for the currents and flow rates in the system, derived in sections \ref{sec:streaming} and \ref{sec:eom}, can be succinctly represented in matrix form as 
\begin{equation}\label{eq:matrixiq}
\begin{pmatrix}
J_{net} \\ Q_{net}
\end{pmatrix} =\begin{pmatrix} L_{\phi} &L_p \\ S_{\phi} &S_p \end{pmatrix} 
\begin{pmatrix} \Delta\phi \\ \Delta P \end{pmatrix}.
\end{equation}
In the above representation, $L_{\phi}=L_{eom}+L_{cond}$, is a sum of the contributions due to the EO current and the bulk conduction current. The coefficients given in equation (\ref{eq:matrixiq}) have been calculated in equations (\ref{eq:istreamfull}), (\ref{eq:ieomfull}), (\ref{eq:iconduction}), (\ref{eq:qp}) and (\ref{eq:qe}). Based on general considerations in non-equilibrium thermodynamics it was shown\cite{mazur,brunet} that the matrix in the above representation is symmetric, i.e. $L_p=S_\phi$, a relation referred to as Onsager reciprocity. We will later see, that this is in fact valid in our approximation and was also obeyed in our numerical simulations. Note that this Onsager relation has been elegantly shown to be valid on striped surfaces by Ng and Chu\citep{Ng_2011} and as such can be used in checking the validity of numerical calculations.

Kirchoff's law dictates that the currents generated in the system be balanced. Since the sign of the EO current is negative, we have
\begin{equation}
\label{eq:currentbalance}
 L_{p}\Delta P+\frac{\Delta\phi}{R_{ext}}+(L_{cond}+ L_{eom})\Delta\phi = 0.
\end{equation}
The streaming potential is given by $\Delta\phi$ when $R_{ext}\rightarrow\infty$. Conversely, when we want to draw maximum power from the circuit, $R_{ext}$ must satisfy $R_{ext}^{-1}=L_{eom}+L_{cond}$. Thus, the maximum power which can be extracted from the circuit is given by
\begin{equation}\label{pmax}
P_{max}=\frac{(\Delta\phi)^2}{R_{ext}}=(\Delta\phi)^2(L_{cond}+L_{eom}).
\end{equation}
The input power into the system is derived from the pressure-driven flow as $P_{in}=\Delta P(Q_p+Q_{e})$. The efficiency of energy conversion, $\eta$, is the ratio between harvested and input power, $\eta=P_{out}/P_{in}$, and is dependent on the external resistance used for extraction. Based on the Onsager reciprocity, the efficiency can be elegantly written in terms of the figure of merit\cite{li}, $Z$,
\begin{equation}\label{eq:alpha}
Z=\frac{S_{\phi}^2}{S_pL_{\phi}},
\end{equation}
and the maximum efficiency then becomes
\begin{equation}\label{eq:efficiency}
\begin{split}
&\eta_{max}=\frac{Z}{(1+\sqrt{1-Z})^2}.
\end{split}
\end{equation}

\section{Results and Discussion}\label{sec:resultsanddiscussion}

Based on the earlier formulations, the effect of electro-osmotic flow on the energy conversion efficiency, output power and the streaming potential developed were calculated. The results are presented in two parts: first we look at the Onsager relation and approximate analytical results; then we look at a more exact graphical representation of the solution. 

\subsection{Approximate analytical solutions}\label{sec:onsager}

The Onsager relation predicts that the amount of current generated per unit pressure drop, $L_p$, should be the same as the flow generated per unit voltage drop for the system, $S_\phi$. The coefficients corresponding to the current generated by the pressure field and the flow generated by the potential are given by equations (\ref{eq:istreamsimple}) and (\ref{eq:qe}) respectively. Thus, we see that under the thin bi-layer assumption ($\kappa h\gg1$)
\begin{equation}
L_{p}=S_\phi=wh\left[\frac{\varepsilon \zeta_s \kappa w}{2\mu l}\beta_\parallel+\frac{\varepsilon \overline \zeta}{\mu l}\right],\label{eq:Lp}
\end{equation}
in accordance with the Onsager relation. In the same limit, the diagonal coefficients in equation ($\ref{eq:matrixiq}$) are given by
\begin{align}
L_{eom}&=\left(\frac{\varepsilon^2 \zeta_s^2 \kappa^2 w^2}{2\mu l}\right)\beta_\parallel,\\
L_{cond}&=\frac{\sigma wh}{l}.
\end{align}
For reasonably large values of the free surface fraction one has $\kappa w\beta_\parallel\gg 1$, so the first term in equation (\ref{eq:Lp}) will dominate this expression and the last term may safely be neglected, which we will do in the following. Moreover, instead of using the potentials $\zeta_s$ and $\zeta_{ns}$ on the free and no-slip surfaces, we will mostly refer to the corresponding surface charge densities, $\sigma_s$ and $\sigma_{ns}$, which in the Debye-H\"{u}ckel approximation are related by
\begin{equation}\label{eq:sigma_DH}
\sigma_s=\varepsilon \zeta_s \kappa, \quad \sigma_{ns}=\varepsilon \zeta_{ns} \kappa.
\end{equation}
Substituting these quantities in equations (\ref{pmax}) and (\ref{eq:alpha}), we obtain simplified analytical expressions for the maximum power and figure of merit, $Z$,
\begin{align}\label{eq:poutsimple}
P_{out}&=\frac{\Delta P^2 w^2 h^2 \beta_\parallel}{8\mu l}
\frac{1}{\left(1+2\frac{\mu\sigma h/w}{\sigma_s^2\beta_\parallel}\right)}, \\
\label{eq:chiout}
Z &=\frac{1}{\left(1+\frac{2}{3}\frac{h/w}{\beta_\parallel}\right)\left(1+2\frac{\mu\sigma h/w}{\sigma_s^2\beta_\parallel}\right)}.
\end{align}
From the expression for the figure of merit, $Z$, the efficiency of power extraction can be easily obtained using equation (\ref{eq:efficiency}). In our approximation, since transport on the free surface is completely dominating the system, both the maximum power extracted and the efficiency of extraction are independent of the charge on the no-slip surface and the Debye layer thickness. For large free-surface charges, a saturation behavior is observed, c.f. equations (\ref{eq:poutsimple}) and (\ref{eq:chiout}). This occurs because of the linear dependence of the streaming current on the charge distribution or the free surface charge, while a quadratic dependence exists in the relationship of the EO current on the free surface charge, since the EO flow which drives the current is itself dependent on the surface charge. The efficiency at saturation depends only on the ratio of (effective) length scales, $h/(w\beta_\parallel)$, and its dependence on the free surface fraction $\delta$ is graphically seen in section~\ref{sec:graphs}. In particular, it only depends on a geometric ratio of (effective) length scales, $h/(w\beta_\parallel)$, and the dimensionless group $\mu\sigma h/(\sigma_s^2w\beta_\parallel)$, which is the ratio between charge transport by bulk conduction and electro-osmotic streaming and determines how close the efficiency (\ref{eq:efficiency}) and power can come to their saturation value.

It is a simple matter to calculate the corresponding efficiency for a parallel plate arrangement, where instead of the no slip condition a Navier slip condition, $b\partial_y u = u$, with slip length $b$ is used at the walls. In the limit $h\kappa\gg1$ and $b \kappa\gg1$, the results in this case are completely analogous to equations (\ref{eq:poutsimple}) and (\ref{eq:chiout}) when we make the identification $b=\beta_\parallel w/2$. Note that this is in complete agreement with and directly related to the finding, that for a striped surface $b=\beta_\parallel w/2$ acts as an effective slip length.

Here we again like to stress the fact that these results, equations (\ref{eq:poutsimple}) and (\ref{eq:chiout}), are valid in the limit of small Debye lengths, $\lambda_D\ll w$, for not too large ratios of $\zeta_s/\zeta_{ns}\ll \beta_\parallel w/\lambda_D$ and only for $h\gtrsim w$. Although the result was obtained in the Debye-H\"uckel approximation, $\zeta< k_BT/e$, this assumption is likely less stringent, as long as we interpret $\varepsilon\zeta_s k$ as the surface charge density in all occurrences, since this quantity is more directly related to the driving force at the interface. In section~\ref{sec:graphs}, we will compare these results with the more exact ones based on expressions (\ref{eq:istreamfull}), (\ref{eq:qp}), (\ref{eq:ieomfull}) and (\ref{eq:qe}). Additionally we also compare with numerical calculations, using the nonlinear Poisson-Boltzmann equation instead of relying on the linearised Debye-H\"{u}ckel approximation.

\subsection{Numerical Methods} 
In order to probe the range of validity of our (semi-)analytical results, we use a commercial finite element package, \textsc{Comsol Multiphysics}\cite{Comsol}, to go beyond the linear Debye-H\"{u}ckel approximation, eq. (\ref{eq:governv}). In particular, the electric potential is obtained by solving the nonlinear Poisson--Boltzmann equation appropriate for a symmetric electrolyte with ions of equal mobility,
\begin{equation}
 \nabla^2 V(x,y) = \kappa^2 V_0 \sinh (V/V_0),
\end{equation}
where $V_0=k_BT/e\approx 26$ mV. On the no-slip and no-shear boundaries the charge density is prescribed, corresponding to a Neumann boundary condition
\begin{equation}
-\varepsilon\partial_y V |_{y=0} = \sigma_i,\quad \sigma_i=\left\{ 
\begin{array}{ll}
\sigma_{s}, & 0<x<\delta w/2, \\ 
\sigma_{ns}, & \delta w/2<x<w/2.
\end{array} 
\right.
\end{equation}
We use here the charge density instead of the zeta potential, since it more directly reflects the driving force for the EO flow on the free surface. We remark that the appropriate conversion between charge density and zeta potential at large potentials can, instead of equation (\ref{eq:sigma_DH}) in the linear regime, be approximately achieved via the Gouy-Chapman solution available for a wall with fixed charge density \cite{Kirby_2010}.

The governing equations for pressure driven and EO flow remain the same, equations (\ref{eq:upEq}) and (\ref{eq:ueEq}), with $\rho_e= -\varepsilon \kappa^2 V_0 \sinh (V/V_0)$ as charge density, however. As before, the streaming current is obtained by integrating the product of velocity field and charge density, equations (\ref{eq:istreamdef}) and (\ref{eq:jeomdef}). Similarly, ion concentrations vary according to the Boltzmann equation, $c_i=c_{bulk} \exp(-z_i V/V0)$, where $c_{bulk}$ is the bulk ion concentration and $z_i=\pm 1$ is its valency. Hence, the local conductivity is given by $\sigma(x,y)=\sigma_{bulk}\cosh (V/V0)$, where $\sigma_{bulk}=2 F \mu_E c_{bulk}$ with ion mobility $\mu_E$ and Faraday's constant $F$. Practically, unless stated otherwise, we prescribe certain values of $\lambda_D$ and $\sigma_{bulk}$ and from these the corresponding values of bulk concentration, $c_{bulk}=\varepsilon V_0/(2F\lambda_D^2)$, and mobility, $\mu_E=\sigma_{bulk}\lambda_D^2/(\varepsilon V_0)$, are inferred. The total conduction current thus becomes
\begin{equation}\label{eq:jCondVar}
J_{cond}=2(\Delta\phi/l)\int_0^h\! dx\! \int_0^{w/2}\! dy\, \sigma(x,y),
\end{equation}
replacing equation (24), which is used apart from cases where we explicitly want to highlight the influence of the Debye layer conductivity. Thus all terms of the transfer matrix (\ref{eq:matrixiq}) are numerically known and the efficiency analysis can proceed as before. 

A typical computational grid consists of $N_x\times N_y=100\times 300$ elements. Particular care was taken to refine the grid close to boundaries at $y=0$ in order to capture the steep gradient in electric potential and EO velocity and obtain grid independent results\cite{GridChoice}. A fifth order discretization was used for all the PDE's.

\subsection{Graphical Analysis}\label{sec:graphs}

In this section, we plot the behavior of the analytical expressions for maximal power and efficiency derived in section~\ref{sec:onsager} in some illustrative examples. These results are compared both with the more exact semi-analytic expressions presented in sections~\ref{sec:streaming}, \ref{sec:eom} and \ref{sec:energyeff} as well as with numerical data to asses the influence of a finite Debye-layer thickness. We assume the fluid properties to be those of water, i.e. $\mu=1$ mPas, $\varepsilon=80\cdot 8.85\,10^{-12}$ As/(Vm). Initially, we will consider a low bulk conductivity of $\sigma=0.1\,\upmu$S/cm of highly purified water and thus also a correspondingly large Debye length of $\lambda_D=1\,\upmu$m for the comparison. Note that for our numerical implementation this corresponds to the hypothetical mobility of $\mu_E\simeq 5.43\cdot 10^{-7}\,\mathrm{m}^2$/(Vs) and bulk concentration $c_{bulk}\simeq 10^{-7}$ mol/L.

\begin{figure}[!t]
\includegraphics[width=\textwidth]{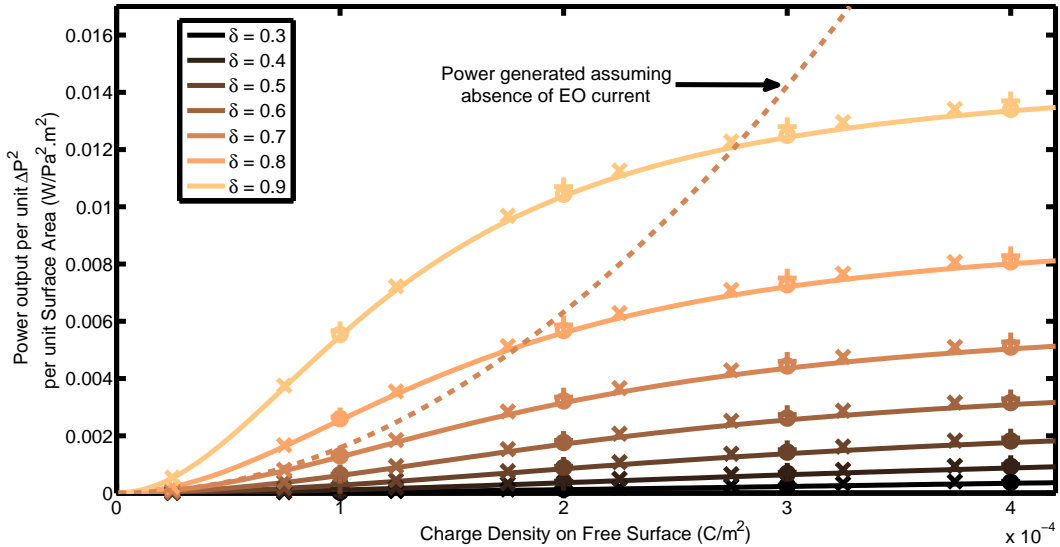}
\caption{Output power density for varying free surface charge density, $\sigma_s$, for different values of the free surface fraction, $\delta$. The solid line is calculated based on our analytical expression (\ref{eq:poutsimple}), and the crosses ($\times$) are calculated based on the more exact semi-analytic expressions (\ref{eq:istreamfull}), (\ref{eq:qp}), (\ref{eq:ieomfull}) and (\ref{eq:qe}). Pluses (+) and circles (o) are numerical results using the nonlinear Poisson-Boltzmann equation, differing in the evaluation of the conduction current by, respectively, a constant and variable charge density according to equations (\ref{eq:iconduction}) and (\ref{eq:jCondVar}). The dotted line shows the behavior when the electro-osmotic flow is not accounted for (at $\delta=0.7$). The figure was plotted assuming $w=h=100\,\upmu$m, $\sigma=0.1\,\upmu$S/cm, $\lambda_D=1\,\upmu$m, $\mu=1$ mPas and $\sigma_s/\sigma_{ns}=2$.}
\label{fig:poutput}
\end{figure} 

Figure~\ref{fig:poutput} shows the variation of the output power with amount of charge on the free surface. With increasing surface charge densities on the free surface the power extracted increases until the electro-osmotic effect begins to dominate. Subsequently, a saturation in the power extracted is achieved. The power output at saturation and the surface charge at which the saturation is reached varies depending on the amount of free surface present, parametrised by the free surface fraction $\delta$. The dashed line in this figure further illustrates the role played by the electro-osmotic flow and its streaming current: when EO flow is neglected, the power extracted from the circuit grows rapidly and without bound with increasing amount of charge on the free surface. Further, our semi-analytic and numerical data suggest that the analytical solution accurately predicts the behavior of the system given a constant in bulk conductivity of the system. In order to highlight the role played by the increase in conductivity within the Debye layer, we have evaluated the conductive contribution to $L_{\phi}$ in the numerical data both assuming a constant charge density, equation (\ref{eq:iconduction}), as assumed for the (semi-) analytic evaluation, as well as with a variable charge density according to equation (\ref{eq:jCondVar}). It is evident that here there is little difference between the two cases.

\begin{figure}[!tb]
\includegraphics[width=\textwidth]{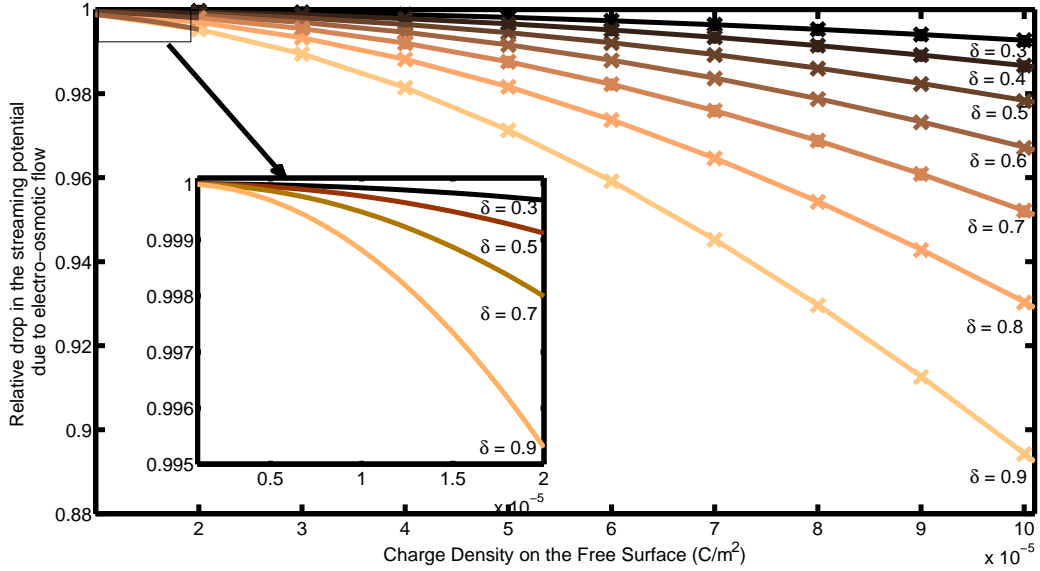}
\caption{Relative drop of streaming potential for increasing free surface charge density, $\sigma_s$, for different values of the free surface fraction, $\delta$. The solid line is calculated based on our analytical expression (\ref{eq:poutsimple}), and the crosses ($\times$) are calculated based on the more exact expressions (\ref{eq:istreamfull}), (\ref{eq:qp}), (\ref{eq:ieomfull}) and (\ref{eq:qe}). The figure was plotted assuming $w=h=100\,\upmu$m, $\sigma=0.1\,\upmu$S/cm, $\lambda_D=1\, \upmu$m, $\mu=1$ mPas and $\sigma_s/\sigma_{ns}=2$.}
\label{fig:relativestreamingdrop}
\end{figure}

The role played by the EO flow and its streaming current is further illustrated in figure~\ref{fig:relativestreamingdrop}, where the relative drop in streaming potential compared to the case without EO flow is shown. We find that the drop in streaming potential when we account for the EO flow is quite significant, especially for higher values of the free surface charge. Further, this drop increases for increasing ratios of the free surface fraction, $\delta$. This increasing contribution of the EO current results in decreasing the streaming potential observed for increasing surface charge. Note that in figure \ref{fig:relativestreamingdrop} compared to figure \ref{fig:poutput} we consider a smaller range of charge densities at the free surface  and that here the difference between analytic and semi-analytic results is negligible.

\begin{figure}[!tb]
\includegraphics[width=\textwidth]{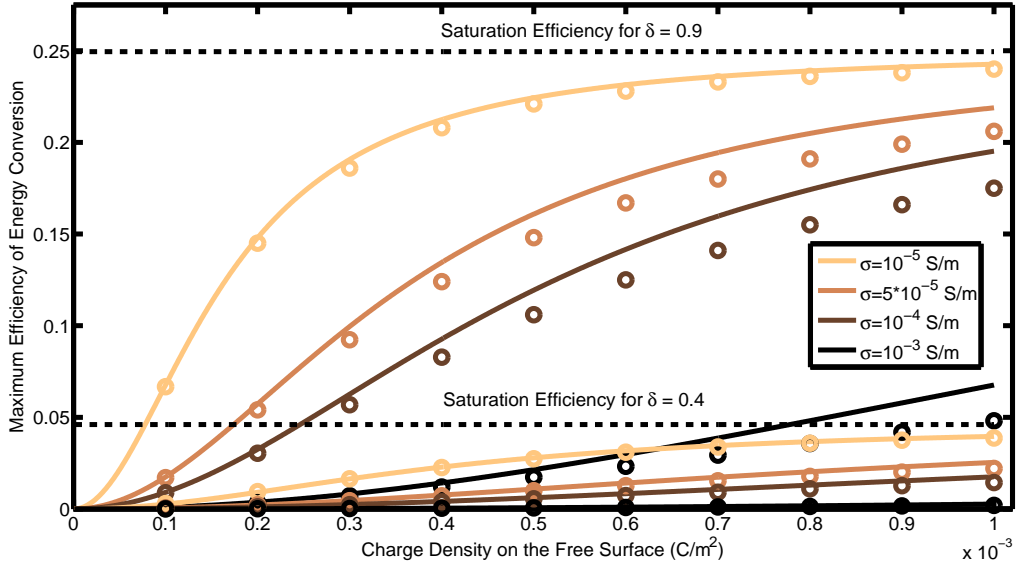}
\caption{Variation of efficiency with free surface charge density, $\sigma_s$, both for varying values of bulk conductivity $\sigma$ and free surface fractions $\delta=0.4$ and $\delta=0.9$. Lines are calculated based on the analytical expressions (\ref{eq:efficiency}, \ref{eq:chiout}). Circles (o) are numerical results obtained using the nonlinear Poisson-Boltzmann equation. The dashed lines shows the maximum theoretical efficiency predicted for such systems. Here $w=h=100\,\upmu$m, $\lambda_D=1\, \upmu$m, $\mu=1$ mPas and $\sigma_s/\sigma_{ns}=2$.
}
\label{fig:efficiencytotal}
\end{figure}

Figure~\ref{fig:efficiencytotal} shows the variation of efficiency for two different values of free surface fraction $\delta$ and for different values of bulk conductivity $\sigma$. We see that the maximum efficiency which can be achieved for these devices is independent of the bulk conductivity of the fluid, as evident from equation (\ref{eq:chiout}) and (\ref{eq:efficiency}); in particular, it is dependent only on the geometry of the channel, such as the width, height, length and the free surface fraction. However, we see that for lower values of $\sigma$, the saturation efficiency is achieved for lower values of the charge density on the free surface. We note that at fixed surface charge density, the efficiency-increase with $\delta$ is not linear and grows more rapidly the closer $\delta$ becomes to 1; a similar trend is followed by the  power output, as shown in figure \ref{fig:poutput}. Again the analytic expression agrees well with the numerical data.

We would like to stress here that, when attracting charges to the free surface by a normal electric field as described in the introduction, a charge density of $10^{-4}$ C/m$^2$ converts to an applied field of $E_{ext}=\sigma_s/\varepsilon_0\approx 10^7$ V/m, about three times the breakdown field strength for dry air. To reach such high surface charge densities by this method, it would thus be necessary to use a medium with higher breakdown voltage or larger permittivity $\varepsilon$.
 (For a discussion of other electric breakdown mechanisms on superhydrophobic surfaces we refer to Oh et al.\cite{Oh_2011}.) Nevertheless, for large enough free surface fractions, efficiencies of the order of a percent seem possible for a microstructured system by this method, something that otherwise is only achieved in nanostructured devices\cite{vdHeyden_2007}.

\begin{figure}[!tb]
\includegraphics[width=\textwidth]{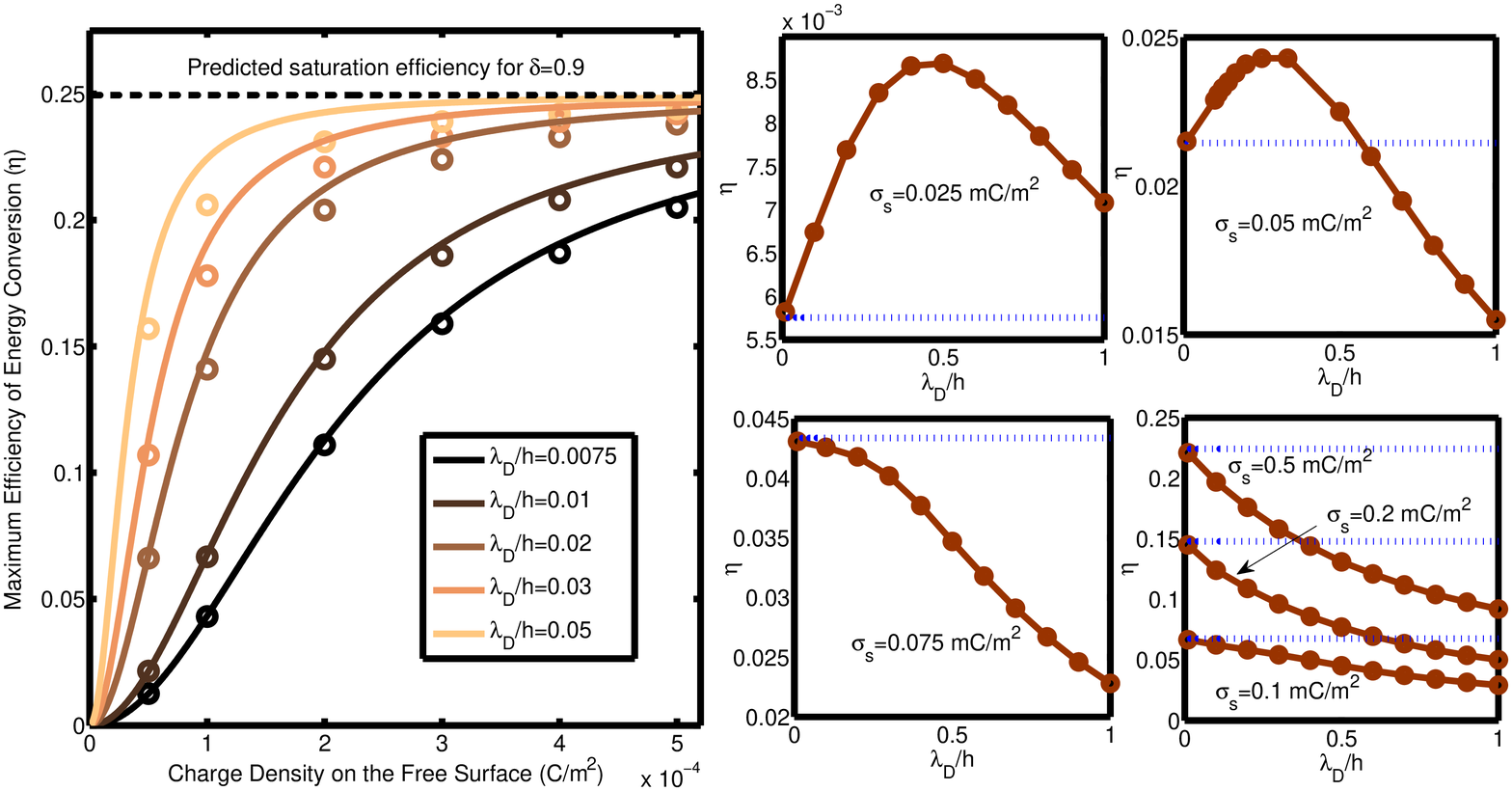}\\
(a)\hspace{7cm}(b)
\caption{(a) Variation of maximum efficiency with free surface charge density, $\sigma_s$, for values of $\lambda_D/h$ between 0.01 and 0.1, fixing the geometry length scale $h$ and varying $\lambda_D$. Lines are calculated based on the analytical expressions (\ref{eq:efficiency}, \ref{eq:chiout}). Circles (o) are numerical results obtained using the nonlinear Poisson-Boltzmann equation. Here $w=h=100\,\upmu$m, $\delta=0.9$, $\mu_E=5.43\cdot 10^{-7}\,\mathrm{m}^2$/(Vs),  $\mu=1$ mPas and $\sigma_s/\sigma_{ns}=2$. (b) Variation of maximum efficiency as a function of $\lambda_D/h$ at fixed $\lambda_D=1\, \upmu$m, $\sigma=0.1\,\upmu$S/cm and variable length-scale $h$ for several values of surface charge density $\sigma_s$. Circles (\textbullet) are numerical results obtained using the nonlinear Poisson-Boltzmann equation and the dotted line is the analytical expression (\ref{eq:efficiency}, \ref{eq:chiout}). Here $w=h$, $\delta=0.9$, $\mu=1$ mPas and $\sigma_s/\sigma_{ns}=2$.
}
\label{fig:varylambda}
\end{figure}

The graphs presented above represent a case where the periodicity of the surface pattern and the height of the channel were assumed to be 100~$\upmu$m, and the fluid properties are based on those of highly purified water. A low conductivity of the system was taken in all cases. In figures \ref{fig:poutput} and \ref{fig:relativestreamingdrop}, the Debye layer thickness was assumed to be a high value of 1$\upmu$m, corresponding to the low concentrations typically present in low conductivity aqueous systems. To a large degree, the behavior of the curves is independent of the Debye Layer thickness and the charge on the no-slip surface, since our analytical expressions conform with the predictions based on our more exact formulations. This slight dependence on the Debye layer thickness disappears for smaller values of $\lambda_D$ and our analytic results, (\ref{eq:poutsimple}) and (\ref{eq:chiout}), become excellent approximations. We find that the efficiency is determined mainly by the ratio of $\beta_\parallel w/h$.  Our results thus indicate that for microfluidic devices with increasingly large portions of free-surface, we would be able to achieve much higher energy conversions due to the steep increase in effective slip length \cite{ybert}.

To gain further insight into the dependence of the efficiency of energy conversion for increasing thickness of the Debye-layer additional numerical calculations were performed. For this we fix the ion mobility, $\mu_E$, and prescribe variable values for $\lambda_D$, in each case choosing the bulk concentration and corresponding conductivity in agreement with these values. In this way we study a fixed electrolyte at varying bulk concentration (and thus varying bulk conductivity). Since we are mainly interested in the impact of $\lambda_D$ on our results and less in a particular electrolyte system we fix the mobility at the hypothetical value of $\mu_E=5.43\cdot 10^{-7}\,\mathrm{m}^2$/(Vs), as used previously. The results of these calculations are found in figure \ref{fig:varylambda} (a) for values of $\lambda_D/w$ varying from 0.0075 to 0.05 and are compared with the analytical expression, (\ref{eq:poutsimple}) and (\ref{eq:chiout}). At fixed surface charge density the efficiency of energy conversion increases for increasing values of the Debye layer thickness. This just reflects the fact that an increasing $\lambda_D$ corresponds to decreasing ion concentrations and thus conductivity. As before the values saturate at the the value predicted by our theoretical expressions for large values of the free surface charge. In addition to that, even below saturation, the analytic expression, equations (\ref{eq:chiout}) and (\ref{eq:efficiency}), approximates the numerical results well. However, it is apparent, that the deviation will become larger with a further increase in Debye length. Nevertheless, this indicates that the validity of our analysis also extends to aqueous systems in smaller structures than presently considered.

To demonstrate this more directly, we also vary $h$ at fixed $\lambda_D$ and bulk conductivity. This is shown in figure \ref{fig:varylambda} (b) for the case of highly purified water, i.e. $\lambda_D=1\, \upmu$m and $\sigma=0.1\,\upmu$S/cm, and for several values of surface charge density $\sigma_s$ between $2.5\cdot 10^{-5}$ and $5\cdot 10^{-4}$C/m$^2$. For small values of the free surface charge the graphs shows the characteristic\cite{davidson} increase of the efficiency with $\lambda_D/h$ to a maximum and the decrease of efficiency upon further reduction of the length scale due to increased average conductivity in the channel. For increasing $\sigma_s$ the maximum is shifted towards smaller values of $\lambda_D/h$, until it vanishes and a monotonic decrease is observed for $\sigma_s \gtrsim 7.5\cdot 10^{-5}$C/m$^2$ in this case. Together with figure \ref{fig:varylambda} (a) this gives a fairly good indication for the range of applicability of the analytic expression, (\ref{eq:poutsimple}) and (\ref{eq:chiout}). Nevertheless, since our main interest lies with larger structures we will not pursue this matter further here.

\section{Conclusion}\label{sec:conclusion}

In this paper we have analyzed the energy conversion from hydraulic pressure to electric energy in a device with large free surface fractions, as observed for a liquid in Cassie-Baxter state on superhydrophobic surfaces. Due to the substantially reduced drag, pressure driven flow is greatly enhanced over such surfaces. Accordingly, when charges are present on these free-slip surfaces one obtains a greatly enhanced streaming current, proportional to the surface charge density. However, in this case also the EO flow and its corresponding streaming current is enhanced by the same mechanism, depending linearly and quadratic on the charge density on the free surface, respectively. Due to this quadratic dependence, the EO streaming current strongly influences and limits the efficiency of the device, leading to a saturation of efficiency for large charge densities on the free surface. The cross-over when EO flow becomes important is dictated by the dimensionless group $\mu\sigma h/(\sigma_s^2w\beta_\parallel)$, the ratio between bulk and EO streaming current. As we have shown, this is equivalent to a device with unstructured surfaces obeying a Navier-slip condition with a slip length of $b=\beta_\parallel w/2$.

This observed saturation with increasing charge density on the free-surface implies that it would be of greater importance to design devices with a higher free surface fraction, which would lead to a considerable increase in the efficiency of energy conversion.  These microstructured devices have the potential to be far more efficient than flat plates with no slip present everywhere and can even reach efficiencies only achieved in nanostructured devices so far. We await further experimental evidence to support the use of such systems both for energy conversion in microfluidic chips and as flow-rate detection devices integrated in conduits or even made for implants on surfaces such as blood vessels.

\begin{acknowledgments}
We thank Mathias Dietzel, Clarissa Sch\"onecker and Steffen Hardt for fruitful discussions. GS kindly acknowledges support by the German Academic Exchange Service (DAAD) through the WISE program. TB kindly acknowledges support by the German Research Foundation (DFG) through the Cluster of Excellence 259.
\end{acknowledgments}

\section*{References}

\begin{thebibliography}{33}%
\makeatletter
\providecommand \@ifxundefined [1]{%
 \@ifx{#1\undefined}
}%
\providecommand \@ifnum [1]{%
 \ifnum #1\expandafter \@firstoftwo
 \else \expandafter \@secondoftwo
 \fi
}%
\providecommand \@ifx [1]{%
 \ifx #1\expandafter \@firstoftwo
 \else \expandafter \@secondoftwo
 \fi
}%
\providecommand \natexlab [1]{#1}%
\providecommand \enquote  [1]{``#1''}%
\providecommand \bibnamefont  [1]{#1}%
\providecommand \bibfnamefont [1]{#1}%
\providecommand \citenamefont [1]{#1}%
\providecommand \href@noop [0]{\@secondoftwo}%
\providecommand \href [0]{\begingroup \@sanitize@url \@href}%
\providecommand \@href[1]{\@@startlink{#1}\@@href}%
\providecommand \@@href[1]{\endgroup#1\@@endlink}%
\providecommand \@sanitize@url [0]{\catcode `\\12\catcode `\$12\catcode
  `\&12\catcode `\#12\catcode `\^12\catcode `\_12\catcode `\%12\relax}%
\providecommand \@@startlink[1]{}%
\providecommand \@@endlink[0]{}%
\providecommand \url  [0]{\begingroup\@sanitize@url \@url }%
\providecommand \@url [1]{\endgroup\@href {#1}{\urlprefix }}%
\providecommand \urlprefix  [0]{URL }%
\providecommand \Eprint [0]{\href }%
\providecommand \doibase [0]{http://dx.doi.org/}%
\providecommand \selectlanguage [0]{\@gobble}%
\providecommand \bibinfo  [0]{\@secondoftwo}%
\providecommand \bibfield  [0]{\@secondoftwo}%
\providecommand \translation [1]{[#1]}%
\providecommand \BibitemOpen [0]{}%
\providecommand \bibitemStop [0]{}%
\providecommand \bibitemNoStop [0]{.\EOS\space}%
\providecommand \EOS [0]{\spacefactor3000\relax}%
\providecommand \BibitemShut  [1]{\csname bibitem#1\endcsname}%
\let\auto@bib@innerbib\@empty
\bibitem [{\citenamefont {Quincke}(1859)}]{quincke}%
  \BibitemOpen
  \bibfield  {author} {\bibinfo {author} {\bibfnamefont {G.}~\bibnamefont
  {Quincke}},\ }\bibfield  {title} {\enquote {\bibinfo {title} {\"{U}ber eine
  neue art elektrischer str\"{o}me},}\ }\href {\doibase
  10.1002/andp.18591830502} {\bibfield  {journal} {\bibinfo  {journal} {Annalen
  der Physik}\ }\textbf {\bibinfo {volume} {183}},\ \bibinfo {pages} {1--47}
  (\bibinfo {year} {1859})}\BibitemShut {NoStop}%
\bibitem [{\citenamefont {Helmholtz}(1879)}]{helmholtz}%
  \BibitemOpen
  \bibfield  {author} {\bibinfo {author} {\bibfnamefont {H.}~\bibnamefont
  {Helmholtz}},\ }\bibfield  {title} {\enquote {\bibinfo {title} {Studien
  \"{u}ber electrische grenzschichten},}\ }\href {\doibase
  10.1002/andp.18792430702} {\bibfield  {journal} {\bibinfo  {journal} {Annalen
  der Physik}\ }\textbf {\bibinfo {volume} {243}},\ \bibinfo {pages} {337--382}
  (\bibinfo {year} {1879})}\BibitemShut {NoStop}%
\bibitem [{\citenamefont {Probstein}(2003)}]{probstein}%
  \BibitemOpen
  \bibfield  {author} {\bibinfo {author} {\bibfnamefont {R.~F.}\ \bibnamefont
  {Probstein}},\ }\href@noop {} {\emph {\bibinfo {title} {Physicochemical
  Hydrodynamics}}}\ (\bibinfo  {publisher} {John Wiley and Sons, Hoboken, New
  Jersey},\ \bibinfo {year} {2003})\BibitemShut {NoStop}%
\bibitem [{\citenamefont {Xuan}\ and\ \citenamefont {Li}(2006)}]{li}%
  \BibitemOpen
  \bibfield  {author} {\bibinfo {author} {\bibfnamefont {X.}~\bibnamefont
  {Xuan}}\ and\ \bibinfo {author} {\bibfnamefont {D.}~\bibnamefont {Li}},\
  }\bibfield  {title} {\enquote {\bibinfo {title} {Thermodynamic analysis of
  electrokinetic energy conversion},}\ }\href {\doibase
  10.1016/j.jpowsour.2005.05.057} {\bibfield  {journal} {\bibinfo  {journal}
  {Journal of Power Sources}\ }\textbf {\bibinfo {volume} {156}},\ \bibinfo
  {pages} {677--684} (\bibinfo {year} {2006})}\BibitemShut {NoStop}%
\bibitem [{\citenamefont {Zhao}(2011)}]{zhao11}%
  \BibitemOpen
  \bibfield  {author} {\bibinfo {author} {\bibfnamefont {H.}~\bibnamefont
  {Zhao}},\ }\bibfield  {title} {\enquote {\bibinfo {title} {Streaming
  potential generated by a pressure-driven flow over superhydrophobic
  stripes},}\ }\href {\doibase 10.1063/1.3551616} {\bibfield  {journal}
  {\bibinfo  {journal} {Phys. Fluids}\ }\textbf {\bibinfo {volume} {23}},\
  \bibinfo {pages} {022003} (\bibinfo {year} {2011})}\BibitemShut {NoStop}%
\bibitem [{\citenamefont {van~der Heyden}\ \emph {et~al.}(2007)\citenamefont
  {van~der Heyden}, \citenamefont {Bonthuis}, \citenamefont {Stein},
  \citenamefont {Meyer},\ and\ \citenamefont {Dekker}}]{vdHeyden_2007}%
  \BibitemOpen
  \bibfield  {author} {\bibinfo {author} {\bibfnamefont {F.~H.~J.}\
  \bibnamefont {van~der Heyden}}, \bibinfo {author} {\bibfnamefont {D.~J.}\
  \bibnamefont {Bonthuis}}, \bibinfo {author} {\bibfnamefont {D.}~\bibnamefont
  {Stein}}, \bibinfo {author} {\bibfnamefont {C.}~\bibnamefont {Meyer}}, \ and\
  \bibinfo {author} {\bibfnamefont {C.}~\bibnamefont {Dekker}},\ }\bibfield
  {title} {\enquote {\bibinfo {title} {Power generation by pressure-driven
  transport of ions in nanofluidic channels},}\ }\href {\doibase
  10.1021/nl070194h} {\bibfield  {journal} {\bibinfo  {journal} {Nano Letters}\
  }\textbf {\bibinfo {volume} {7}},\ \bibinfo {pages} {1022--1025} (\bibinfo
  {year} {2007})}\BibitemShut {NoStop}%
\bibitem [{\citenamefont {van~der Heyden}\ \emph {et~al.}(2006)\citenamefont
  {van~der Heyden}, \citenamefont {Bonthuis}, \citenamefont {Stein},
  \citenamefont {Meyer},\ and\ \citenamefont {Dekker}}]{vdHeyden_2006}%
  \BibitemOpen
  \bibfield  {author} {\bibinfo {author} {\bibfnamefont {F.~H.~J.}\
  \bibnamefont {van~der Heyden}}, \bibinfo {author} {\bibfnamefont {D.~J.}\
  \bibnamefont {Bonthuis}}, \bibinfo {author} {\bibfnamefont {D.}~\bibnamefont
  {Stein}}, \bibinfo {author} {\bibfnamefont {C.}~\bibnamefont {Meyer}}, \ and\
  \bibinfo {author} {\bibfnamefont {C.}~\bibnamefont {Dekker}},\ }\bibfield
  {title} {\enquote {\bibinfo {title} {Electrokinetic energy conversion
  efficiency in nanofluidic channels},}\ }\href {\doibase 10.1021/nl061524l}
  {\bibfield  {journal} {\bibinfo  {journal} {Nano Letters}\ }\textbf {\bibinfo
  {volume} {6}},\ \bibinfo {pages} {2232--2237} (\bibinfo {year}
  {2006})}\BibitemShut {NoStop}%
\bibitem [{\citenamefont {Yang}\ and\ \citenamefont {Kwok}(2003)}]{yang_2003}%
  \BibitemOpen
  \bibfield  {author} {\bibinfo {author} {\bibfnamefont {J.}~\bibnamefont
  {Yang}}\ and\ \bibinfo {author} {\bibfnamefont {D.~Y.}\ \bibnamefont
  {Kwok}},\ }\bibfield  {title} {\enquote {\bibinfo {title} {Microfluid flow in
  circular microchannel with electrokinetic effect and navier's slip
  condition},}\ }\href {\doibase 10.1021/la026201t} {\bibfield  {journal}
  {\bibinfo  {journal} {Langmuir}\ }\textbf {\bibinfo {volume} {19}},\ \bibinfo
  {pages} {1047--1053} (\bibinfo {year} {2003})}\BibitemShut {NoStop}%
\bibitem [{\citenamefont {Yang}\ and\ \citenamefont {Kwok}(2004)}]{yang_2004}%
  \BibitemOpen
  \bibfield  {author} {\bibinfo {author} {\bibfnamefont {J.}~\bibnamefont
  {Yang}}\ and\ \bibinfo {author} {\bibfnamefont {D.~Y.}\ \bibnamefont
  {Kwok}},\ }\bibfield  {title} {\enquote {\bibinfo {title} {Analytical
  treatment of electrokinetic microfluidics in hydrophobic microchannels},}\
  }\href {\doibase 10.1016/j.aca.2003.12.043} {\bibfield  {journal} {\bibinfo
  {journal} {Analytica Chimica Acta}\ }\textbf {\bibinfo {volume} {507}},\
  \bibinfo {pages} {39--53} (\bibinfo {year} {2004})}\BibitemShut {NoStop}%
\bibitem [{\citenamefont {Davidson}\ and\ \citenamefont
  {Xuan}(2008)}]{davidson}%
  \BibitemOpen
  \bibfield  {author} {\bibinfo {author} {\bibfnamefont {C.}~\bibnamefont
  {Davidson}}\ and\ \bibinfo {author} {\bibfnamefont {X.}~\bibnamefont
  {Xuan}},\ }\bibfield  {title} {\enquote {\bibinfo {title} {Electrokinetic
  energy conversion in slip nanochannels},}\ }\href {\doibase
  10.1016/j.jpowsour.2007.12.050} {\bibfield  {journal} {\bibinfo  {journal}
  {Journal of Power Sources}\ }\textbf {\bibinfo {volume} {179}},\ \bibinfo
  {pages} {297--300} (\bibinfo {year} {2008})}\BibitemShut {NoStop}%
\bibitem [{\citenamefont {Ren}\ and\ \citenamefont {Stein}(2008)}]{Ren_2008}%
  \BibitemOpen
  \bibfield  {author} {\bibinfo {author} {\bibfnamefont {Y.}~\bibnamefont
  {Ren}}\ and\ \bibinfo {author} {\bibfnamefont {D.}~\bibnamefont {Stein}},\
  }\bibfield  {title} {\enquote {\bibinfo {title} {Slip-enhanced electrokinetic
  energy conversion in nanofluidic channels},}\ }\href {\doibase
  10.1088/0957-4484/19/19/195707} {\bibfield  {journal} {\bibinfo  {journal}
  {Nanotechnology}\ }\textbf {\bibinfo {volume} {19}},\ \bibinfo {pages}
  {195707} (\bibinfo {year} {2008})}\BibitemShut {NoStop}%
\bibitem [{\citenamefont {Goswami}\ and\ \citenamefont
  {Chakraborty}(2011)}]{chakraborty}%
  \BibitemOpen
  \bibfield  {author} {\bibinfo {author} {\bibfnamefont {P.}~\bibnamefont
  {Goswami}}\ and\ \bibinfo {author} {\bibfnamefont {S.}~\bibnamefont
  {Chakraborty}},\ }\bibfield  {title} {\enquote {\bibinfo {title}
  {Semi-analytical solutions for electroosmotic flows with interfacial slip in
  microchannels of complex cross-sectional shapes},}\ }\href {\doibase
  10.1007/s10404-011-0793-6} {\bibfield  {journal} {\bibinfo  {journal}
  {Microfluidics and Nanofluidics}\ }\textbf {\bibinfo {volume} {11}},\
  \bibinfo {pages} {255--267} (\bibinfo {year} {2011})}\BibitemShut {NoStop}%
\bibitem [{\citenamefont {Ybert}\ \emph {et~al.}(2007)\citenamefont {Ybert},
  \citenamefont {Barentin}, \citenamefont {Cottin-Bizonne}, \citenamefont
  {Joseph},\ and\ \citenamefont {Bocquet}}]{ybert}%
  \BibitemOpen
  \bibfield  {author} {\bibinfo {author} {\bibfnamefont {C.}~\bibnamefont
  {Ybert}}, \bibinfo {author} {\bibfnamefont {C.}~\bibnamefont {Barentin}},
  \bibinfo {author} {\bibfnamefont {C.}~\bibnamefont {Cottin-Bizonne}},
  \bibinfo {author} {\bibfnamefont {P.}~\bibnamefont {Joseph}}, \ and\ \bibinfo
  {author} {\bibfnamefont {L.}~\bibnamefont {Bocquet}},\ }\bibfield  {title}
  {\enquote {\bibinfo {title} {Achieving large slip with superhydrophobic
  surfaces: Scaling laws for generic geometries},}\ }\href {\doibase
  10.1063/1.2815730} {\bibfield  {journal} {\bibinfo  {journal} {Physics of
  Fluids}\ }\textbf {\bibinfo {volume} {19}},\ \bibinfo {eid} {123601}
  (\bibinfo {year} {2007})}\BibitemShut {NoStop}%
\bibitem [{\citenamefont {Rothstein}(2010)}]{rothstein}%
  \BibitemOpen
  \bibfield  {author} {\bibinfo {author} {\bibfnamefont {J.~P.}\ \bibnamefont
  {Rothstein}},\ }\bibfield  {title} {\enquote {\bibinfo {title} {Slip on
  superhydrophobic surfaces},}\ }\href {\doibase
  10.1146/annurev-fluid-121108-145558} {\bibfield  {journal} {\bibinfo
  {journal} {Annual Review of Fluid Mechanics}\ }\textbf {\bibinfo {volume}
  {42}},\ \bibinfo {pages} {89--109} (\bibinfo {year} {2010})}\BibitemShut
  {NoStop}%
\bibitem [{\citenamefont {Philip}(1972)}]{philip}%
  \BibitemOpen
  \bibfield  {author} {\bibinfo {author} {\bibfnamefont {J.~R.}\ \bibnamefont
  {Philip}},\ }\bibfield  {title} {\enquote {\bibinfo {title} {Flows satisfying
  mixed no-slip and no-shear conditions},}\ }\href {\doibase
  10.1007/BF01595477} {\bibfield  {journal} {\bibinfo  {journal} {J. Applied
  Mathematics and Physics}\ }\textbf {\bibinfo {volume} {23}},\ \bibinfo
  {pages} {353--372} (\bibinfo {year} {1972})}\BibitemShut {NoStop}%
\bibitem [{\citenamefont {Lauga}\ and\ \citenamefont {Stone}(2003)}]{lauga}%
  \BibitemOpen
  \bibfield  {author} {\bibinfo {author} {\bibfnamefont {E.}~\bibnamefont
  {Lauga}}\ and\ \bibinfo {author} {\bibfnamefont {H.}~\bibnamefont {Stone}},\
  }\bibfield  {title} {\enquote {\bibinfo {title} {Effective slip in
  pressure-driven stokes flow.}}\ }\href {\doibase 10.1017/S0022112003004695}
  {\bibfield  {journal} {\bibinfo  {journal} {Journal of Fluid Mechanics}\
  }\textbf {\bibinfo {volume} {489}},\ \bibinfo {pages} {55--77} (\bibinfo
  {year} {2003})}\BibitemShut {NoStop}%
\bibitem [{\citenamefont {Sbragaglia}\ and\ \citenamefont
  {Prosperetti}(2007)}]{prosperetti}%
  \BibitemOpen
  \bibfield  {author} {\bibinfo {author} {\bibfnamefont {M.}~\bibnamefont
  {Sbragaglia}}\ and\ \bibinfo {author} {\bibfnamefont {A.}~\bibnamefont
  {Prosperetti}},\ }\bibfield  {title} {\enquote {\bibinfo {title} {A note on
  the effective slip properties for microchannel flows with ultrahydrophobic
  surfaces},}\ }\href {\doibase 10.1063/1.2716438} {\bibfield  {journal}
  {\bibinfo  {journal} {Phys. Fluids}\ }\textbf {\bibinfo {volume} {19}},\
  \bibinfo {pages} {043603} (\bibinfo {year} {2007})}\BibitemShut {NoStop}%
\bibitem [{\citenamefont {Bazant}\ and\ \citenamefont
  {Vinogradova}(2008)}]{Bazant_2008}%
  \BibitemOpen
  \bibfield  {author} {\bibinfo {author} {\bibfnamefont {M.}~\bibnamefont
  {Bazant}}\ and\ \bibinfo {author} {\bibfnamefont {O.}~\bibnamefont
  {Vinogradova}},\ }\bibfield  {title} {\enquote {\bibinfo {title} {Tensorial
  hydrodynamic slip},}\ }\href {\doibase 10.1017/S002211200800356X} {\bibfield
  {journal} {\bibinfo  {journal} {J. Fluid Mech.}\ }\textbf {\bibinfo {volume}
  {613}},\ \bibinfo {pages} {125--134} (\bibinfo {year} {2008})}\BibitemShut
  {NoStop}%
\bibitem [{\citenamefont {David M.~Huang}\ and\ \citenamefont
  {Bocquet}(2008)}]{Huang_2008}%
  \BibitemOpen
  \bibfield  {author} {\bibinfo {author} {\bibfnamefont {C.~Y.}\ \bibnamefont
  {David M.~Huang}, \bibfnamefont {C\'{e}cile Cottin-Bizonne}}\ and\ \bibinfo
  {author} {\bibfnamefont {L.}~\bibnamefont {Bocquet}},\ }\bibfield  {title}
  {\enquote {\bibinfo {title} {Massive amplification of surface-induced
  transport at superhydrophobic surfaces},}\ }\href {\doibase
  10.1103/PhysRevLett.101.064503} {\bibfield  {journal} {\bibinfo  {journal}
  {Phys. Rev. Lett.}\ }\textbf {\bibinfo {volume} {101}},\ \bibinfo {pages}
  {064503} (\bibinfo {year} {2008})}\BibitemShut {NoStop}%
\bibitem [{\citenamefont {Melcher}\ and\ \citenamefont
  {Taylor}(1969)}]{melcher}%
  \BibitemOpen
  \bibfield  {author} {\bibinfo {author} {\bibfnamefont {J.~R.}\ \bibnamefont
  {Melcher}}\ and\ \bibinfo {author} {\bibfnamefont {G.~I.}\ \bibnamefont
  {Taylor}},\ }\bibfield  {title} {\enquote {\bibinfo {title}
  {Electrohydrodynamics: A review of the role of interfacial shear stresses},}\
  }\href {\doibase 10.1146/annurev.fl.01.010169.000551} {\bibfield  {journal}
  {\bibinfo  {journal} {Annual Review of Fluid Mechanics}\ }\textbf {\bibinfo
  {volume} {1}},\ \bibinfo {pages} {111--146} (\bibinfo {year}
  {1969})}\BibitemShut {NoStop}%
\bibitem [{\citenamefont {Steffes}, \citenamefont {Baier},\ and\ \citenamefont
  {Hardt}(2011)}]{steffes}%
  \BibitemOpen
  \bibfield  {author} {\bibinfo {author} {\bibfnamefont {C.}~\bibnamefont
  {Steffes}}, \bibinfo {author} {\bibfnamefont {T.}~\bibnamefont {Baier}}, \
  and\ \bibinfo {author} {\bibfnamefont {S.}~\bibnamefont {Hardt}},\ }\bibfield
   {title} {\enquote {\bibinfo {title} {Enabling the enhancement of
  electroosmotic flow over superhydrophobic surfaces by induced charges},}\
  }\href {\doibase 10.1016/j.colsurfa.2010.09.002} {\bibfield  {journal}
  {\bibinfo  {journal} {Colloids and Surfaces A: Physicochemical and
  Engineering Aspects}\ }\textbf {\bibinfo {volume} {376}},\ \bibinfo {pages}
  {85--88} (\bibinfo {year} {2011})}\BibitemShut {NoStop}%
\bibitem [{\citenamefont {Morrison}\ and\ \citenamefont
  {Osterle}(1965)}]{osterle}%
  \BibitemOpen
  \bibfield  {author} {\bibinfo {author} {\bibfnamefont {F.~A.}\ \bibnamefont
  {Morrison}}\ and\ \bibinfo {author} {\bibfnamefont {J.~F.}\ \bibnamefont
  {Osterle}},\ }\bibfield  {title} {\enquote {\bibinfo {title} {Electrokinetic
  energy conversion in ultrafine capillaries},}\ }\href {\doibase
  10.1063/1.1697081} {\bibfield  {journal} {\bibinfo  {journal} {Journal of
  Chemical Physics}\ }\textbf {\bibinfo {volume} {43}},\ \bibinfo {pages}
  {2111} (\bibinfo {year} {1965})}\BibitemShut {NoStop}%
\bibitem [{\citenamefont {Squires}(2008)}]{squires}%
  \BibitemOpen
  \bibfield  {author} {\bibinfo {author} {\bibfnamefont {T.~M.}\ \bibnamefont
  {Squires}},\ }\bibfield  {title} {\enquote {\bibinfo {title} {Electrokinetic
  flows over inhomogeneously slipping surfaces},}\ }\href {\doibase
  10.1063/1.2978954} {\bibfield  {journal} {\bibinfo  {journal} {Physics of
  Fluids}\ }\textbf {\bibinfo {volume} {20}},\ \bibinfo {eid} {092105}
  (\bibinfo {year} {2008})}\BibitemShut {NoStop}%
\bibitem [{\citenamefont {Bahga}, \citenamefont {Vinogradova},\ and\
  \citenamefont {Bazant}(2010)}]{bazant1}%
  \BibitemOpen
  \bibfield  {author} {\bibinfo {author} {\bibfnamefont {S.~S.}\ \bibnamefont
  {Bahga}}, \bibinfo {author} {\bibfnamefont {O.~I.}\ \bibnamefont
  {Vinogradova}}, \ and\ \bibinfo {author} {\bibfnamefont {M.~Z.}\ \bibnamefont
  {Bazant}},\ }\bibfield  {title} {\enquote {\bibinfo {title} {Anisotropic
  electro-osmotic flow over super-hydrophobic surfaces},}\ }\href {\doibase
  10.1017/S0022112009992771} {\bibfield  {journal} {\bibinfo  {journal} {J.
  Fluid Mech.}\ }\textbf {\bibinfo {volume} {644}},\ \bibinfo {pages}
  {245--255} (\bibinfo {year} {2010})}\BibitemShut {NoStop}%
\bibitem [{\citenamefont {Belyaev}\ and\ \citenamefont
  {Vinogradova}(2011)}]{Belyaev_2011}%
  \BibitemOpen
  \bibfield  {author} {\bibinfo {author} {\bibfnamefont {A.}~\bibnamefont
  {Belyaev}}\ and\ \bibinfo {author} {\bibfnamefont {O.}~\bibnamefont
  {Vinogradova}},\ }\bibfield  {title} {\enquote {\bibinfo {title}
  {Electro-osmosis on anisotropic super-hydrophobic surfaces},}\ }\href
  {\doibase 10.1103/PhysRevLett.107.098301} {\bibfield  {journal} {\bibinfo
  {journal} {Phys. Rev. Lett.}\ }\textbf {\bibinfo {volume} {107}},\ \bibinfo
  {pages} {098301} (\bibinfo {year} {2011})}\BibitemShut {NoStop}%
\bibitem [{\citenamefont {Ng}\ and\ \citenamefont {Chu}(2011)}]{Ng_2011}%
  \BibitemOpen
  \bibfield  {author} {\bibinfo {author} {\bibfnamefont {C.-O.}\ \bibnamefont
  {Ng}}\ and\ \bibinfo {author} {\bibfnamefont {H.~C.~W.}\ \bibnamefont
  {Chu}},\ }\bibfield  {title} {\enquote {\bibinfo {title} {Electrokinetic
  flows through a parallel-plate channel with slipping stripes on walls},}\
  }\href {\doibase 10.1063/1.3647582} {\bibfield  {journal} {\bibinfo
  {journal} {Phys. Fluids}\ }\textbf {\bibinfo {volume} {23}},\ \bibinfo
  {pages} {102002} (\bibinfo {year} {2011})}\BibitemShut {NoStop}%
\bibitem [{\citenamefont {Sneddon}(1966)}]{sneddon}%
  \BibitemOpen
  \bibfield  {author} {\bibinfo {author} {\bibfnamefont {I.~N.}\ \bibnamefont
  {Sneddon}},\ }\href@noop {} {\emph {\bibinfo {title} {Mixed Boundary Value
  Problems in Potential Theory}}}\ (\bibinfo  {publisher} {North-Holland,
  Amsterdam},\ \bibinfo {year} {1966})\BibitemShut {NoStop}%
\bibitem [{\citenamefont {Mazur}\ and\ \citenamefont {Overbeek}(1951)}]{mazur}%
  \BibitemOpen
  \bibfield  {author} {\bibinfo {author} {\bibfnamefont {P.}~\bibnamefont
  {Mazur}}\ and\ \bibinfo {author} {\bibfnamefont {J.~T.~G.}\ \bibnamefont
  {Overbeek}},\ }\bibfield  {title} {\enquote {\bibinfo {title} {On
  electro-osmosis and streaming-potentials in diaphragms: Ii. general
  quantitative relationship between electro-kinetic effects},}\ }\href
  {\doibase 10.1002/recl.19510700114} {\bibfield  {journal} {\bibinfo
  {journal} {Recueil des Travaux Chimiques des Pays-Bas}\ }\textbf {\bibinfo
  {volume} {70}},\ \bibinfo {pages} {83--91} (\bibinfo {year}
  {1951})}\BibitemShut {NoStop}%
\bibitem [{\citenamefont {Brunet}\ and\ \citenamefont {Ajdari}(2004)}]{brunet}%
  \BibitemOpen
  \bibfield  {author} {\bibinfo {author} {\bibfnamefont {E.}~\bibnamefont
  {Brunet}}\ and\ \bibinfo {author} {\bibfnamefont {A.}~\bibnamefont
  {Ajdari}},\ }\bibfield  {title} {\enquote {\bibinfo {title} {Generalized
  onsager relations for electrokinetic effects in anisotropic and heterogeneous
  geometries},}\ }\href {\doibase 10.1103/PhysRevE.69.016306} {\bibfield
  {journal} {\bibinfo  {journal} {Phys. Rev. E}\ }\textbf {\bibinfo {volume}
  {69}},\ \bibinfo {pages} {016306} (\bibinfo {year} {2004})}\BibitemShut
  {NoStop}%
\bibitem [{Com()}]{Comsol}%
  \BibitemOpen
  \href@noop {} {}\bibinfo {note} {COMSOL Multiphysics v4.2, COMSOL AB,
  Stockholm, Sweden}\BibitemShut {NoStop}%
\bibitem [{\citenamefont {Kirby}(2010)}]{Kirby_2010}%
  \BibitemOpen
  \bibfield  {author} {\bibinfo {author} {\bibfnamefont {B.}~\bibnamefont
  {Kirby}},\ }\href@noop {} {\emph {\bibinfo {title} {Micro- and Nanoscale
  Fluid Mechanics}}}\ (\bibinfo  {publisher} {Cambridge University Press},\
  \bibinfo {year} {2010})\BibitemShut {NoStop}%
\bibitem [{Gri()}]{GridChoice}%
  \BibitemOpen
  \href@noop {} {}\bibinfo {note} {Useful guides for the choice of grid are
  comparisons in limiting cases with known analytic solutions such as the
  Gouy-Chapman potential distribution\cite{Kirby_2010} at a uniformly charged
  wall and the corresponding charge density and EO velocity, Philip's
  solution\cite{philip} for the velocity field over a surface with slip/no-slip
  stripes driven by a constant stress on the free surface (corresponding to EO
  flow in the limit of infinitely thin charge layer), and the validity of
  Onsager's relations.}\BibitemShut {Stop}%
\bibitem [{\citenamefont {Oh}\ \emph {et~al.}(2011)\citenamefont {Oh},
  \citenamefont {Manukyan}, \citenamefont {van~den Ende},\ and\ \citenamefont
  {Mugele}}]{Oh_2011}%
  \BibitemOpen
  \bibfield  {author} {\bibinfo {author} {\bibfnamefont {J.~M.}\ \bibnamefont
  {Oh}}, \bibinfo {author} {\bibfnamefont {G.}~\bibnamefont {Manukyan}},
  \bibinfo {author} {\bibfnamefont {D.}~\bibnamefont {van~den Ende}}, \ and\
  \bibinfo {author} {\bibfnamefont {F.}~\bibnamefont {Mugele}},\ }\bibfield
  {title} {\enquote {\bibinfo {title} {Electric-field–-driven instabilities
  on superhydrophobic surfaces},}\ }\href {\doibase 10.1209/0295-5075/93/56001}
  {\bibfield  {journal} {\bibinfo  {journal} {EPL (Europhysics Letters)}\
  }\textbf {\bibinfo {volume} {93}},\ \bibinfo {pages} {56001} (\bibinfo {year}
  {2011})}\BibitemShut {NoStop}%
\end{thebibliography}

%

\newpage
\appendix
\section{Simplifying the Electro-Osmotic Current}\label{app:eomsimplify}
The EO streaming current, equation (\ref{eq:jeomdef}), can be simplified along the same lines as the streaming current due to pressure driven flow in equations (\ref{eq:istreamfull})-(\ref{eq:istreamsimple}).
Using the definition of the potential distribution provided in equation (\ref{eq:vsplit}) and the expressions for the electro-osmotic velocity given by equations (\ref{eq:uesplit}) and (\ref{eq:uecorrection}), we simplify equation (\ref{eq:jeomdef}) to get
\begin{equation}
\label{ieomanalytic}
\begin{split}
J_{eom}&=2(-\varepsilon \kappa^2)[\int_0^h\int_0^\frac{\delta w}{2} (\zeta_se^{-\kappa y})(\frac{\zeta_s\varepsilon\Delta\phi}{\mu l})(1-e^{-\kappa y})+\int_0^h\int_\frac{\delta w}{2}^\frac{w}{2} (\zeta_{ns}e^{-\kappa y})(\frac{\zeta_{ns}\varepsilon\Delta\phi}{\mu l})(1-e^{-\kappa y})\\&+
         \int_0^h\int_0^\frac{w}{2}\left[\overline\zeta e^{-\kappa y}+\sum_{n=1}^\infty \beta_n\cos\left(\frac{2n\pi x}{w}\right)e^{-\sqrt{\kappa^2+(\frac{2n\pi}{w})^2}y}\right]\left[A_0+\sum_{n=1}^\infty A_n\cos\left(\frac{2n\pi x}{w}\right)e^{-\frac{2n\pi y}{w}} \right]
\end{split}
\end{equation}
Performing the integrations in equation (\ref{ieomanalytic}), leads to
\begin{equation}
\label{ieomapprox}
\begin{split}
J_{eom}=2(-\varepsilon \kappa^2)\Bigg[\frac{\zeta_s^2\varepsilon\Delta\phi\delta w}{4\kappa\mu l}+\frac{\zeta_{ns}^2\varepsilon\Delta\phi(1-\delta) w}{4\kappa\mu l}+\frac{\overline\zeta w^2\zeta_s\varepsilon\Delta\phi}{2\pi\mu l}\log\sec(\frac{\delta\pi}{2})\\
+\frac{(\zeta_s-\zeta_{ns})w^2}{2}\sum_{n=1}^\infty \frac{A_n\sin(n\pi d)}{n\pi(2n\pi+\sqrt{(\kappa w)^2+(2n\pi)^2}}\Bigg].
\end{split}
\end{equation}
Assuming that $\kappa w\gg 1$, we can approximate equation (\ref{ieomapprox}) as
\begin{equation}
\label{eq:eom_simple}
\begin{split}
J_{eom}=2(-\varepsilon \kappa^2)\Bigg[\frac{\zeta_s^2\varepsilon\Delta\phi\delta w}{4\kappa\mu l}+\frac{\zeta_{ns}^2\varepsilon\Delta\phi(1-\delta) w}{4\kappa\mu l}+\frac{\overline\zeta w^2\zeta_s\varepsilon\Delta\phi}{2\pi\mu l}\log\sec\left(\frac{\delta\pi}{2}\right)\\
+ \frac{(\zeta_s-\zeta_{ns})w}{2\kappa}\sum_{n=1}^\infty \frac{A_n\sin(n\pi d)}{n\pi}\Bigg].
\end{split}
\end{equation}
By integrating equation (\ref{eq:a0}) from 0 to $\frac{\delta\pi}{2}$, we can show that $\sum_{n=1}^\infty \frac{A_n\sin(n\pi d)}{n\pi}=(1-\delta)A_0$. Therefore, equation (\ref{eq:eom_simple}) can be written as
\begin{equation}
J_{eom}=2(-\varepsilon \kappa^2)\left(\frac{\zeta_s^2\varepsilon\Delta\phi w}{4\kappa\mu l}\right)\left(d+\left(\frac{\zeta_s}{\zeta_{ns}}\right)^2(1-d)+\frac{2w\kappa}{\pi}\log\sec\frac{\delta\pi}{2}\right).
\end{equation}
For reasonably large values of $\delta$ where $\log\sec\frac{\delta\pi}{2}\simeq O(1)$, we see that because $w\kappa\gg 1$ we have $\frac{2w\kappa}{\pi}\log\sec\frac{\delta\pi}{2}\gg d+(\frac{\zeta_s}{\zeta{ns}})^2(1-d)$. Therefore, $J_{eom}$ can be approximately written as
\begin{equation}
J_{eom}=2(-\varepsilon \kappa^2)\left(\frac{\zeta_s^2\varepsilon\Delta\phi w}{4\kappa\mu l}\right)\left(\frac{2w\kappa}{\pi}\log\sec\frac{\delta\pi}{2}\right).
\end{equation}

\end{document}